\documentclass[twocolumn,showpacs,superscriptaddress,pra]{revtex4}

\usepackage{amsmath, amsthm, amssymb}
\usepackage{graphicx}
\usepackage{dcolumn}
\usepackage{bm}
\usepackage{color}

\newcommand{\ket}[1]{| #1 \rangle}
\newcommand{\bra}[1]{\langle #1 |}

\begin{document}
\title{Multipartite quantum nonlocality under local decoherence}
\author{R. Chaves}
\affiliation{ICFO-Institut de Ci\`encies Fot\`oniques, Mediterranean
Technology Park, 08860 Castelldefels (Barcelona), Spain}
\affiliation{Instituto de F\'\i sica, Universidade Federal do Rio
de Janeiro. Caixa Postal 68528, 21941-972 Rio de Janeiro, RJ,
Brasil}
\author{D. Cavalcanti}
\affiliation{Centre for Quantum Technologies, University of
Singapore, Singapore}
\author{L. Aolita}
\affiliation{ICFO-Institut de Ci\`encies Fot\`oniques, Mediterranean
Technology Park, 08860 Castelldefels (Barcelona), Spain}
\author{A. Ac\'in}
\affiliation{ICFO-Institut de Ci\`encies Fot\`oniques, Mediterranean
Technology Park, 08860 Castelldefels (Barcelona), Spain}
\affiliation{ICREA-Instituci\'o Catalana de Recerca i Estudis
Avan\c cats, Lluis Companys 23, 08010 Barcelona, Spain}

\begin{abstract}
We study the nonlocal properties of two-qubit maximally-entangled and $N$-qubit Greenberger-Horne-Zeilinger states  under local decoherence. We show that the
(non)resilience of entanglement under local depolarization or dephasing is
not necessarily equivalent to the (non)resilience of Bell-inequality
violations. Apart from entanglement and Bell-inequality
violations, we consider also nonlocality as quantified by the nonlocal content of correlations, and  provide several examples of anomalous behaviors, both in the bipartite and multipartite cases. In addition, we study the practical implications of these anomalies on the usefulness of noisy Greenberger-Horne-Zeilinger states as resources for  nonlocality-based physical protocols given by communication complexity
problems. There, we provide examples of quantum gains improving with the number of particles that coexist with exponentially-decaying entanglement and nonlocal contents.
\end{abstract}

\maketitle
\section{Introduction}

Although closely connected, entanglement and nonlocality
constitute two substantially different concepts. Entanglement
refers to whether a state can be decomposed as a convex combination of product
quantum states and is therefore inherent to the Hilbert-space
structure of quantum theory~\cite{horod}. Operationally, a state
is entangled whenever it cannot be prepared by local quantum
operations and classical communication. Nonlocality on the other
hand refers to correlations between distant measurements
--whatever the underlying theory-- that cannot be explained by
local hidden-variable models~\cite{Bell}.

Correlations describable in terms of local hidden variables
necessarily satisfy a set of linear constraints known as {\it Bell
inequalities}~\cite{Bell}, which can be tested in the lab. Thus,
the violation of any Bell inequality reveals the presence of
nonlocality; whereas its nonviolation does not have any
implication on the local or nonlocal nature of the corresponding
correlations (unless all Bell inequalities are proven to be
satisfied). In turn, every pure quantum state is entangled if, and
only if, it violates some Bell inequality~\cite{gisin91pr92}.
Additionally, if an arbitrary quantum state is nonlocal, it is
also entangled~\cite{Werner}. The converse however has long been
known not to be true:  There exist mixed entangled states that
admit local hidden-variable descriptions~\cite{Werner,BarretAcin}.

From an applied point of view, entanglement has been identified
over the last two decades as the key resource in a variety of
physical tasks (see Ref.~\cite{horod} and references therein).
These go from teleportation, dense coding, secure quantum
key distribution (QKD) and quantum communication, to quantum
computing, for instance. In more recent years,  it was
nonlocality that was also raised to the status of a physical
resource. An example thereof was given by the advent of
device-independent applications, such as QKD~\cite{DIQKD} or
randomness generation~\cite{randgen}. There, correlations violating
some Bell inequality suffice to establish a secret key or a perfectly random bit, regardless of the physical means by which they are established. Another prominent example is
distributed-computing scenarios such as those of {\it communication
complexity problems} (CCPs)~\cite{Buhrman,CCPBrukner}. There, $N$
distant users, assisted by some initial correlations and  a
restricted amount of public communication, must locally calculate
the value of a given function $f$ with some probability of
success. It was shown in Ref.~\cite{CCPBrukner} that, for a broad
family of $N$-partite Bell inequalities, one can associate to
every inequality a CCP that can be solved more efficiently (with
higher probability) than by any classical protocol if, and only
if, it is assisted by correlations that violate the inequality.
Furthermore, the authors showed that,  for any fixed $N$, the quantum
gain (in success probability) is proportional to the amount of
violation, thus automatically yielding a direct operational
interpretation for the violation of this type of Bell
inequality.

However,  under realistic  situations where actual
applications take place, systems are unavoidably subject to noise. It is therefore important to probe the resilience of physical
resources in the presence of noise. This becomes particularly necessary for
many-particle systems, where the detrimental effects of the
interaction with the environment typically accumulate
exponentially with the number of system components. Nevertheless,
while the open-system dynamics of multipartite entanglement has
been extensively studied
\cite{SimonKempe&Carvalho&HeinDurBrie&Guehne&Wunderlich, Aolita,
Graphdeco,Jack}, the scaling behavior of the nonlocal properties
of quantum states under decoherence is barely understood. To the
best of our knowledge,  some cases of such behavior were systematically
studied in Ref.~\cite{noisyviolations}, but focused only on the
critical noise strengths (or, alternatively, the times) for which
the violations of a specific family of multisetting $N$-partite
Bell inequalities vanish. We know, however, from the study of
entanglement, that such critical values on their own can be very
misleading as figures of merit of any robustness. Situations are
known where correlations take longer to vanish but still, for a
given fixed time, decay exponentially with $N$ \cite{Aolita}.
Thus, the full dynamical evolution of nonlocality must  be studied
to draw conclusions on its fragility.

In this paper we study
the evolution of nonlocality, in comparison to that of
entanglement, for two-qubit maximally entangled and Greenberger-Horne-Zeilinger
(GHZ)~\cite{GHZ} states subject to independent depolarization or
phase damping~\cite{Nie}. We find that
decoherence can lead to unexpected behavior in the nonlocal
properties of states \cite{anomaly}. For instance, in the two-qubit case, local decoherence can lead to the natural appearance of anomalies in the orderings of states, such that the less entangled states have more nonlocality. In turn, in the multipartite case, we identify regimes of noise for which
the violation of the Mermin inequality ~\cite{Mermin} by decohered GHZ states
grows exponentially with the number of particles $N$, despite the fact that both entanglement and nonlocal content \cite{EPR2} decay exponentially. Remarkably, in some cases, this exponentially decaying entanglement is even bound \cite{Aolita}. To get a physical insight of the practical consequences of such anomalous behavior, we study the quantum gains for CCPs. We find that the gains increase with $N$ for small $N$ and decrease with $N$ for large $N$. Thus, the initial regime of increasing quantum gain coexists with exponentially decaying entanglement and nonlocal content.

\section{Notation and definitions}

In this section we introduce the definitions, concepts, and tools
applied in the derivation of the results.

\subsection{Multi-qubit states under noisy channels}
The initial
states we will consider throughout are  the $GHZ$
states~\cite{GHZ}
\begin{equation}
\left\vert \Phi^+\right\rangle =\frac{1}{\sqrt{2}}(\left\vert 0\right\rangle ^{\otimes
N}+\left\vert 1\right\rangle ^{\otimes N}), \label{GHZ}%
\end{equation}
which reduce to maximally entangled Bell states for $N=2$.

As paradigmatic models of noise we focus on the independent
depolarizing (D) and phase-damping (PD) channels \cite{Nie}. Channel D describes the situation in
which a qubit remains untouched with probability $1-p$ or is mapped
to the maximally mixed state (white noise) with probability $p$:
\begin{equation}\label{depchannel}
    \varepsilon^D_p(\rho)=(1-p)\rho+p\frac{1}{2} .
\end{equation}
Channel PD induces the complete loss of quantum coherence with
probability $p$, but without any population exchange,
\begin{equation}\label{dephchannel}
    \varepsilon^{PD}_p(\rho)=(1-p)\rho+p\sum_i\ket{i}\!\bra{i}\bra{i}\rho\ket{i}
    ,
\end{equation}
where $i$ denotes the energy basis. Probability $p$ can also be interpreted as a convenient
parametrization of time, where $p=0$ refers to the initial time $t=0$
and $p=1$ refers to the asymptotic limit $t\rightarrow\infty$.

The action of these models of noise on states \eqref{GHZ} has been
explicitly calculated several times before. Here, we  make
use of the expressions derived in Ref. \cite{Aolita} and refer the
interested reader to Refs.
\cite{SimonKempe&Carvalho&HeinDurBrie&Guehne&Wunderlich,Aolita} for more details.
For example, states~\eqref{GHZ} under independent depolarizing or phase-damping
channels can in both cases be expressed
as \cite{Aolita}
\begin{equation}
\rho_{(p)} = (1-p)^{N} \ket{\Phi^+}\bra{\Phi^+}
+ \left( 1- (1-p)^{N} \right) \rho_{s},
\label{GHZdecohered}
\end{equation}
where $\rho_{s}$ is a separable
state, diagonal in the computational basis
$\{\ket{0...0},\ket{0...01}, ... ,
\ket{1...1}\}$,
that depends on the channel in question.

The simplicity of decomposition (\ref{GHZdecohered}) allows for
exhaustive analytical treatments. For example, the entanglement
$E(\rho_{(p)})$ -- as quantified by any convex entanglement measure
$E$ -- of decohered states \eqref{GHZdecohered} always decays faster than exponentially with $N$:
$E(\rho_{(p)})\leq(1-p)^{N}E(\rho_{(0)})$~\cite{Aolita}. In
particular, for small  $p$ and large  $N$, it typically saturates
the inequality as $E(\rho_{(p)})\approx(1-p)^{N}E(\rho_{(0)})$.
Exponentially small physical perturbations are enough to fully
disentangle $\rho_{(p)}$. Furthermore, as also mentioned, for the
case of channel D and for any noise strength, there always exists
an $N$ from which on the entanglement in states
\eqref{GHZdecohered} is bound~\cite{Aolita}. We will see in what
follows how the above-mentioned symmetry can also be exploited to
understand the nonlocal properties of the states.

\subsection{Bell-inequality violations by noisy multi-qubit states}
 We consider throughout the same type of Bell inequalities as considered
in Ref. \cite{CCPBrukner}:
\begin{equation}
\mathcal{I}_{N}\doteq{\displaystyle\sum\limits_{x_{1}\ ...\ x_{N}=0}^{1}}
g(x_{1}, ...,x_{N}) C(x_{1}, ... , x_{N})\leq\mathcal{I}^L_{N}. \label{Bellineq}%
\end{equation}
Each part $i$ measures randomly
in one of two settings, $x_{i}=0$ or $x_{i}=1$, and obtains  $1$
or $-1$ as the outcome. Here, $g(x_{1}, ... , x_{N})$ is any real-valued function and
$C(x_{1}, ... , x_{N})$ denotes the correlation function for the
measurements of $N$ separated parties.  $\mathcal{I}^L_{N}$ is in
turn the local bound, that is, the maximum possible value of
polynomial $\mathcal{I}_{N}$ attainable by any
local-hidden-variable (LHV) model. In the quantum case, $x_{i}=0$ or $x_{i}=1$ correspond to
observables $O_{i_{0}}$ or $O_{i_{1}}$, respectively, each one
with eigenvalues $\pm 1$. Then, the correlation function is
given by  $C(x_{1}, ... ,
x_{N})=\text{Tr}[\rho\cdot O_{1_{x_1}}\otimes\ ...\otimes
O_{N_{x_N}}]$, where $\rho$ is the state under scrutiny.

\subsection{Nonlocal content of noisy multi-qubit states}
We will see next
that, apart from the entanglement, the nonlocality of
$\varrho_{(p)}$ also decays as the state's violation grows
exponentially. For this, we consider a measure of nonlocality
based on the Elitzur-Popescu-Rohrlich (EPR2) decomposition
\cite{EPR2}: Any  joint-probability distribution $P$, which
characterizes the correlations of some experiment, can be
decomposed into the convex mixture of purely local and purely
nonlocal parts as
\begin{equation}
\label{EPR2Decomp}
P = (1-p_{NL}) P_{L} + p_{NL}P_{NL}.
\end{equation}
$P_{L}$ and $P_{NL}$ are respectively the corresponding local and nonsignalling
joint-probability distributions in the decomposition, and $0\leq p_{NL}\leq1$. The minimal weight of the
nonlocal part over all such possible decompositions provides an
unambiguous quantification of the nonlocality in $P$:
\begin{equation}
\label{ptilde}
\tilde{p}_{NL}\doteq\min_{P_{L},P_{NL}}p_{NL}.
\end{equation}
It is also called the {\it nonlocal content of} $P$ and its counterpart
$\tilde{p}_{L}\equiv1-\tilde{p}_{NL}$ gives the {\it local
content} of $P$.

The violation of any Bell inequality allows one to obtain a nontrivial lower bound to the nonlocal content. Indeed, for any
(linear) Bell inequality $\mathcal{I}\leq\mathcal{I}^L$, the
optimal decomposition $P= (1- \tilde{p}_{NL}) \tilde{P}_{L} +
\tilde{p}_{NL}\tilde{P}_{NL}$ yields $\mathcal{I}(P)\equiv(1-
\tilde{p}_{NL}) \mathcal{I}(\tilde{P}_{L})+
\tilde{p}_{NL}\mathcal{I}(\tilde{P}_{NL})$. Now, on the one hand,
since it is  local, $\tilde{P}_{L}$ cannot violate the inequality:
$\mathcal{I}(\tilde{P}_{L})\leq\mathcal{I}^L$. On the other hand
$\mathcal{I}(\tilde{P}_{NL})$ cannot be larger than the maximal nonsignalling value, $\mathcal{I}^{NL}$, of $\mathcal{I}$. Therefore,
it is  always $\mathcal{I}(P)\leq(1- \tilde{p}_{NL})\mathcal{I}^L+
\tilde{p}_{NL}\mathcal{I}^{NL}$, from which follows that
\begin{equation}
\label{lowerbound}
    \tilde{p}_{NL}\geq
\frac{\mathcal{I}(P)-\mathcal{I}^{L}}{\mathcal{I}^{NL}-\mathcal{I}^{L}}.
\end{equation}
Notice that any correlations $P$ that violate a Bell inequality
saturating the maximal nonsignalling value,
$\mathcal{I}(P)=\mathcal{I}^{NL}$ are automatically fully
nonlocal (i.e., with  $\tilde{p}_{NL}=1$). This is precisely what
happens to states \eqref{GHZ}, which saturate the algebraic
violations of an infinite-setting Bell inequality
\cite{BarretKentPironio} for $N=2$, and that of the Mermin
inequality \cite{Mermin} for an odd number of parties. This is why states \eqref{GHZ} are said to be maximally
nonlocal. Actually, GHZ states are maximally genuine multipartite
nonlocal, as they reach the algebraic violation of a Bell
inequality for this form of nonlocality~\cite{aolita}.

\section{Results}

\subsection{Anomalies in the noisy dynamics of nonlocality versus entanglement}

\begin{figure} [!t]
\centering
\includegraphics[width=\linewidth]{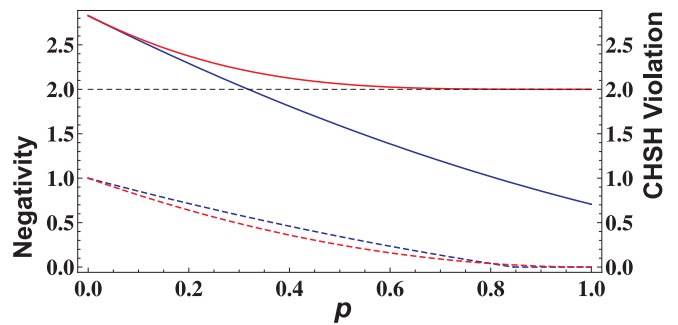}
\caption {(Color online.) Entanglement (lower dashed curves) and maximal value of the CHSH polynomial $\mathcal{I}_{CHSH}$ (upper solid curves), for maximally entangled two-qubit states under independent phase damping of noise strength $p$ (in red, upper solid and lower dashed curves) and independent depolarization of noise strength $p/2$ (in blue, middle two curves), as a function of $p$. The horizontal black dashed line represents the local bound $\mathcal{I}^L_{CHSH}=2$, below which there is no more CHSH violation.  Decoherence naturally drives the system to dephased states that possess less entanglement than depolarized states but that, at the same time, violate the CHSH inequality  more. } \label{fig:anomaly}
\end{figure}

The first example we consider is the familiar Clauser-Horne-Shimony-Holt (CHSH) inequality~\cite{CHSH}, $\mathcal{I}_{2}\equiv\mathcal{I}_{CHSH}$,
defined by Eq. \eqref{Bellineq} (for $N=2$) with $g(x_{1},
x_{2})\equiv (-1)^{x_{1} x_{2}}$ and
$\mathcal{I}^L_{2}\equiv\mathcal{I}^L_{CHSH}=2$. Its maximal
quantum violation ($\mathcal{I}_{CHSH}=2\sqrt{2}$) is realized by
Bell state \eqref{GHZ} (for $N=2$) with the observables
$O_{1_{0}}=-X_1$, $O_{1_{1}}=Z_1$,
$O_{2_{0}}=\frac{X_2-Z_2}{\sqrt{2}}$ and
$O_{2_{1}}=\frac{X_2+Z_2}{\sqrt{2}}$, where  $X_i$ and $Z_i$ are
respectively the first and third Pauli operators of qubit $i$. In
the noisy scenario, the maximal violation for decohered states
\eqref{GHZdecohered} is immediately calculated with the criterion
of Ref. \cite{HoroCHSH}. In Fig.~\ref{fig:anomaly} we
compare the evolution of the maximal value of $\mathcal{I}_{CHSH}$
with that of entanglement,  as a function of $p$, for the cases of
independent phase damping of noise strength $p$ and independent
depolarization of noise strength $p/2$. Entanglement is quantified
by the negativity \cite{Negativity}, which --since the considered states are
Bell diagonal-- coincides with the concurrence \cite{Concu} and can be taken as an unambiguous entanglement quantifier. A curious effect is observed: While the depolarized states display
more entanglement than the dephased ones (except for large $p$),
the violation of the CHSH inequality given by the dephased states
is always above that of the depolarized states. In fact,  as $p$
increases from zero to the point where the depolarized states stop
violating the inequality, the gap between the entanglement and
the violation grows.

\textbf{Result 0:} {\it Local phase damping can naturally drive two-qubit systems to states with less entanglement but more CHSH violation than those driven by local depolarization}.

\begin{figure} [!t]
\centering
\includegraphics[width=\linewidth]{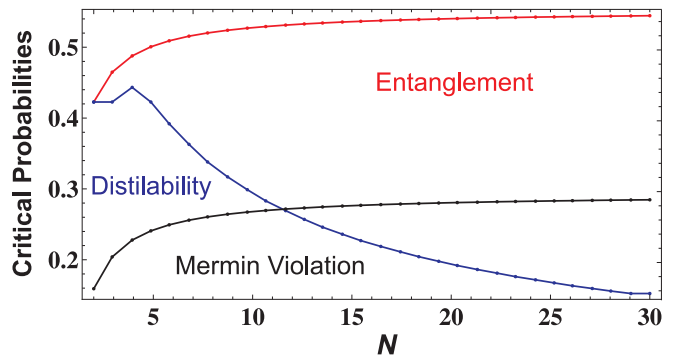}
\caption{(Color online) Critical probabilities for full-separability (in
red), distillability (in blue) and violation of the Mermin
inequality ($p_c$, in black), for $N$-qubit GHZ states under
independent depolarization of strength $p$. Below the red curve
the states are entangled,  they are distillable though only below
the blue curve~\cite{Aolita}. Between the red and the blue curves
the states are thus bound entangled. The total entanglement (not
plotted) decreases exponentially with $N$. Below  the black curve
in turn the states violate the Mermin inequality and they do it
exponentially with $N$. Hence, for $N>11$, and between the black
and the blue curves, decoherence naturally drives the system to
states with exponentially-small and bound entanglement that yet
violate the inequality exponentially.} \label{fig:comparison}
\end{figure}

As $N$ increases this type of anomaly becomes stronger. For $N>2$
we consider the Mermin inequality \cite{Mermin}, defined by
inequality \eqref{Bellineq} with $g(x_{1}, ... ,
x_{N})\equiv\cos\left(\frac{\pi}{2}\left(x_{1}+\ldots+x_{N}\right)\right)$
and $\mathcal{I}^L_{N}=2^{N/2}$, for $N$ even, and
$\mathcal{I}^L_{N}=2^{\left( N-1\right)/2}$, for $N$ odd. Its
maximal quantum violation is $\mathcal{I}_{N}=2^{N-1}$ and is
attained by GHZ states \eqref{GHZ} (for $N>2$) with observables
$O_{i_{0}}=X_i$ and $O_{i_{1}}=Y_i$, where $X_i$ and $Y_i$ are, respectively, the first and second
Pauli operators acting on qubit $i$. Since the local bound
$\mathcal{I}^L_{N}$ is also an exponentially-growing function of
$N$, instead of $\mathcal{I}_{N}$ one usually quantifies the
violation with the ratio
$\mathcal{V}_{N}=\mathcal{I}_{N}/\mathcal{I}^L_{N}$.
$\mathcal{V}_{N}$ is sometimes called the visibility of the
inequality and, in terms of it, inequality \eqref{Bellineq} reads
$\mathcal{V}_{N}\leq1$. It is immediate to see that the maximal
violation of the Mermin inequality for noisy GHZ states
\eqref{GHZdecohered} takes place with the same observables as for
$p=0$. Thus,  their maximal visibility is immediately  calculated  to be
\begin{equation}
\mathcal{V}_{N}\mathcal{=}\left( 1-p\right) ^{N}  2^{\left(N-1\right) /2},
\label{NoisyVis}
\end{equation}
where for simplicity we have taken $N$ odd. Another curious effect
appears here. The entanglement in $\varrho_{(p)}$ decays always
smoothly (exponentially) with $N$ (except for channel D and when
$p\gtrsim0.49$)  \cite{Aolita}. Nevertheless, its visibility
displays an abrupt transition from exponentially growing to
exponentially decreasing at the relatively small value
$p_{\text{t}}=1-1/\sqrt{2}\approx 0.29$, for the two channels
considered. In turn, the critical noise strength beyond which the
inequality is not violated any further is
$p_{\text{c}}=1-1/\sqrt{2}^{\left( N-1\right) /N}<p_{\text{t}}$.
This leads to the following effect: For  $p<p_{\text{c}}$, and as $N$
increases, $\varrho_{(p)}$ gets exponentially close to the
separable states while at the same time yielding an exponentially
large violation of Eq. \eqref{Bellineq}. Furthermore, we know that, for channel D, the little entanglement remaining in
$\varrho_{(p)}$  very rapidly becomes bound \cite{Aolita}. The
situation is illustrated in Fig. \ref{fig:comparison}, where the
regions with the different regimes  are plotted.

\textbf{Result 1:} {\it States with exponentially-small and bound entanglement can provide an
exponentially large Bell inequality violation}.

In addition, from Eqs. \eqref{NoisyVis} and \eqref{lowerbound},  we obtain
\begin{equation}
\label{lowerp}
    \tilde{p}_{NL}\geq \frac{(1-p)^{N}2^{( N-1)/2}-1}{2^{(
N-1)/2}-1} ,
\end{equation}
where again for simplicity we have taken $N$ odd. Additionally,
decomposition \eqref{GHZdecohered} of $\rho_{(p)}$ immediately
yields an upper bound to its local content. This is because the
correlations $P(\ket{\Phi^+})$ in $\ket{\Phi^+}$ are purely
nonlocal whereas correlations $P(\rho_{s})$ of $\rho_{s}$ are purely local (because
$\rho_{s}$ is separable). Therefore,  $P(\rho_{(p)})=(1-p)^{N}P(\ket{\Phi^+})+\left( 1- (1-p)^{N}
\right)P(\rho_{s})$ realizes a particular EPR2 decomposition of
the correlations $P(\rho_{(p)})$ in $\rho_{(p)}$, and  the optimal
one must thus necessarily satisfy
\begin{equation}
\label{upperp}
\tilde{p}_{NL}\leq(1-p)^{N},
\end{equation}
 in a way reminiscent to GHZ entanglement decay \cite{Aolita}.

\textbf{Result 2:} {\it The nonlocal content
in locally depolarized, or dephased, states \eqref{GHZdecohered} cannot decay slower than
exponentially with $N$}.

Notice further that, for $p<p_{\text{t}}=1-1/\sqrt{2}$, lower   bound \eqref{lowerp} converges, as $N$ grows, to the upper bound  \eqref{upperp}. Then,  in the limit
$N\rightarrow\infty$,  the exact value for the nonlocal content
of $\rho_{(p)}$ is $\tilde{p}_{NL}=(1-p)^{N}$. So, exponentially large Mermin visibility \eqref{NoisyVis} coexists not only with an exponentially small (and in some cases bound) entanglement but also with an exponentially small nonlocal content. This indicates that the visibility of a Bell inequality may not always constitute an unambiguous quantitative figure of merit for the nonlocal resources of quantum states.

\subsection{Efficiency gain in CCPs with noisy multiqubit states}

As discussed, the previous results may be a manifestation of the fact that the visibility of a Bell test does not necessarily quantify a state's usefulness for a practical (nonlocal) problem. In particular, in view of both the entanglement and nonlocal content of $\rho_{(p)}$ decaying at slowest exponentially with $N$, it is interesting to explore what implication the observed exponentially growing Mermin-inequality visibility has on some concrete physical task. Here, we focus on the gain in efficiency --with respect to all protocols assisted by classical
correlations-- for solving probabilistic distributed computations given by communication complexity
problems (CCPs) \cite{Buhrman}. In the pure-state case of $p=0$, an exponentially growing Mermin visibility is responsible for an exponentially growing quantum gain \cite{CCPBrukner}. However, for $p>0$, we find that the visibility in question yields a  gain that, after a short transient period of growth with $N$, converges to the universal exponential-decay law of $(1-p)^{N}$.

\begin{figure} [!h]
\centering
\includegraphics[width=0.9\linewidth]{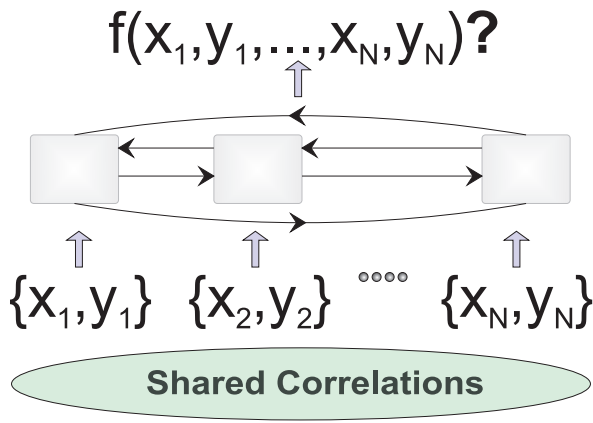}
\caption{(Color online)  Distributed computing scenario \cite{Buhrman}. $N$
distant users receive each a two-bit input string $\{x_{i},
y_{i}\}$, with $1\leq i\leq N$. Distributed bits $y_{i}$ are
chosen randomly between $1$ and $-1$, and each $x_{i}$ is chosen
as $0$ or $1$ depending on a joint probability distribution
$Q(x_{1}, ... , x_{N})$. The users  are endowed with some
pre-established shared correlations, but they can only exchange  a
restricted amount of public communication. The problem is, for each
and all of them, to compute the value
$f(x_{1},...,x_{n},y_{1},...,y_{n})$ of a given function $f$ with
some probability.  The minimum number of bits that they must
broadcast to do so defines the communicational complexity of the
problem. We consider here a specific sub-class of such problems
with $Q(x_{1}, ... , x_{N})\equiv \frac{|g(x_{1},
...,x_{N})|}{\sum_{x_{1}, ...,x_{N}=0}^1|g(x_{1}, ...,x_{N})|}$,
for $g(x_{1}, ...,x_{N})$ some real-valued function, $f$ a boolean
function of the form $f=y_{1}\times ...\ y_{n}\times S[g(
x_{1},...,x_{n})]$, with $f=\pm1$ and $S\left[ g\right]\equiv
g/\left\vert g\right\vert =\pm1$ the sign function, and where each
user is allowed to broadcast only a single bit. For this subclass,
it was shown in Ref. \cite{CCPBrukner},  for a broad family of
protocols,  that if the shared correlations violate Bell inequality
\eqref{Bellineq} the users can solve the problem with a higher
probability than with any classical protocol (assisted by LHV
correlations). \label{fig:CCP}}
\end{figure}
\par The family of CCPs we consider is described in Fig. \ref{fig:CCP}.
For these, the maximal probability of success, $p^s$, achievable
through a broad class of protocols \cite{CCPBrukner} with
pre-established correlations $P$ as the resource is
\begin{equation}
    p^s= \frac{1}{2}\Big(1+\frac{\mathcal{I}_{N}(P)}{\sum_{x_{1},
...,x_{N}=0}^1\left\vert g(x_{1}, ...,x_{N}) \right\vert}\Big) .
\end{equation}
Therefore, no such strategy can do as well when using classical
resources as when based on correlations that violate inequality
\eqref{Bellineq}. For the Mermin inequality one has
$\mathcal{I}^L_{N}=2^{N/2}$ for $N$ even, $\mathcal{I}^L_{N}=2^{(
N-1)/2}$ for $N$ odd, and $\sum_{x_{1}, ...,x_{N}=0}^1\left\vert
g(x_{1}, ...,x_{N})\right\vert=2^{N-1}$. Therefore,  the best such
protocol with  classical correlations solves the associated CCP
with a probability $p^s_L =\frac{1}{2}\big(
1+1/\sqrt{2^{N-1}}\big)$, for $N$ odd, and $p^s_L
=\frac{1}{2}\big( 1+1/\sqrt{2^{N-2}}\big)$, for $N$ even. If
states \eqref{GHZdecohered} are used as the resource in contrast
the protocol succeeds with  $p^s_Q=\frac{1}{2}\big(
1+(1-p)^{N}\big)$.

\par
\begin{figure} [!t]
\centering
\includegraphics[width=\linewidth]{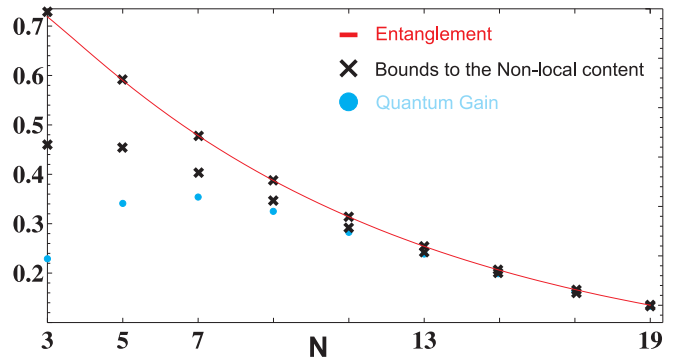}
\caption {
(Color online) Quantum gain $G_Q(p,N)$ --for odd $N$-- (blue circles)
in the efficiency to solve CCPs with $\varrho(p)$, lower and upper
bounds --also for odd $N$-- to the nonlocal content
$\tilde{p}_{NL}$ in $\varrho(p)$ (black crosses), and negativity (solid red) in
the most robust bipartitions of $\varrho(p)$, for $p=0.1$ and as a
function of $N$. For $3\leq N\leq 7$ the quantum gain displays a region of growth. However, for large $N$, it very rapidly converges to
the universal upper bound $(1-p)^N$, and so do all other curves.} \label{quantumgain}
\end{figure}

 The quantum gain in the protocol is defined as $G_{Q}\doteq
p^s_Q-p^s_L$. For $N$ odd, for instance,  it reads
\begin{equation}
\label{QGEq}
G_Q\equiv
G_Q(p,N)=\frac{1}{2}\Big((1-p)^{N}-1/\sqrt{2^{N-1}}\Big).
\end{equation}
From this, we can see that, for $p=0$, $G_Q$ grows monotonically and exponentially with $N$, converging to the maximal value $1/2$ in the limit of $N\to\infty$. For $p>0$, however, both terms in Eq. \eqref{QGEq} decay exponentially and their difference displays a nonmonotonic behavior, with a transition from growing to decaying with $N$ at $N\approx\frac{\log\big(\sqrt{2}\ln(1/\sqrt{2})/\ln(1-p)\big)}{\log\big(\sqrt{2}(1-p)\big)}$. Quantum gain \eqref{QGEq} is plotted in Fig. \ref{quantumgain} as a function
of $N$ for $p=0.1$. Together with it, in the figure, also the lower \cite{comment} and upper bounds
-- for $N$ odd-- to the nonlocal content in $\varrho(p)$, as
well as the negativity in its half-versus-half bipartitions, are plotted. These
bipartitions are the most robust ones and the disappearance (red
upper curve of Fig. \ref{fig:comparison}) of their negativity
characterizes the full separability of $\varrho(p)$ \cite{Aolita}. Therefore, the plotted negativity can be taken as a valid quantitative
figure of merit for the total entanglement of $\varrho(p)$. In the figure, we can see how, after a short region of growth, from $N=3$ to $N=7$, $G_Q$ becomes a decreasing function of $N$. Furthermore, we can see how it rapidly converges to the universal upper bound $(1-p)^{N}$, as well as
all other curves in the figure for large $N$. This constitutes our final result.

\textbf{Result 3:} {\it The usefulness for the nonlocal  task associated with the Mermin inequality of locally depolarized, or dephased, states \eqref{GHZdecohered} decays, for large $N$,
exponentially with $N$}.

\par So an exponentially large visibility renders, apart from coexisting already with small  entanglement and nonlocal content, an exponentially small usefulness for solving the associated physical problem. We emphasize that this remarkable anomaly can happen only in a noisy scenario, as for pure states a monotonically growing visibility implies a monotonically growing gain. Nevertheless, it constitutes a further confirmation of the fact that, as discussed below Result 2, at present day we still do not possess fully satisfactory tools for the unambiguous quantification of the entanglement or nonlocal resources of quantum states.

\section{Conclusion.}

In this work we have analyzed how local noisy environments affect
the nonlocal properties of $N$-qubit states. Interestingly, the
derived picture is more complex than initially expected, as there
are regimes where, although entanglement and nonlocal content show an
exponential decay with the number of parties, the violation of
some Bell inequalities exponentially increase with $N$. This
improvement in the Bell violation may in fact be the reason for
the existence of a regime of $N$ in which decaying entanglement
and nonlocal content coexists with an increasing efficiency to solve
probabilistic distributed computing tasks. Our results, then,
provide a new manifestation of the subtle relation between
entanglement and nonlocality in quantum states. We hope that our
work opens new perspectives in the study of nonlocality decay
under decoherence. In particular, it would be interesting to study
how the recent proposals for robust encoding of $N$-qubit
entanglement introduced in Refs.~\cite{dur,robust} apply to
nonlocality.

\begin{acknowledgements}
This work is supported by the the EU Q-Essence project, the ERC
Starting Grant PERCENT, the Chist-Era DIQIP project, the Spanish
FIS2010-14830 project and Juan de la Cierva foundation, the Brazilian agency FAPERJ and the
National Institute of Science and Technology for Quantum
Information, the National Research Foundation and the Ministry of
Education of Singapore.

\end{acknowledgements}


\begin{thebibliography}{99}
\bibitem{comment} In general the exact value of $\tilde{p}_{NL}$ for a quantum state
is the weight of the nonlocal parts of the optimal EPR2
decompositions maximized over all possible measurements on the
state. While for a given setting, the optimal nonlocal weight can
be efficiently obtained via linear programming, the measurement
optimization is a hard computational task. For  states \eqref{GHZdecohered} of  $N=3$ and $5$, we maximized the
nonlocal content restricted to the correlations over all projective measurements with two settings and two outcomes, and assuming the same measurement settings for each part.
A perfect coincidence, up to numerical precision, with the lower
bound was observed.
\end{thebibliography}
\end{document}